\documentclass[pdflatex,sn-mathphys-num]{sn-jnl}

\usepackage{graphicx}%
\usepackage{multirow}%
\usepackage{amsmath,amssymb,amsfonts}%
\usepackage{amsthm}%
\usepackage{mathrsfs}%
\usepackage[title]{appendix}%
\usepackage{xcolor}%
\usepackage{textcomp}%
\usepackage{manyfoot}%
\usepackage{booktabs}%
\usepackage{algorithm}%
\usepackage{algorithmicx}%
\usepackage{algpseudocode}%
\usepackage{listings}%
\usepackage{bm}
\usepackage{subfigure}

\newcommand{\btheta}{\bm{\theta}} 
\newcommand{\bmu}{\bm{\mu}} 
\newcommand{\bSigma}{\bm{\Sigma}} 
\newcommand{\bTheta}{\bm{\Theta}} 
\newcommand{\bmeta}{\bm{\eta}} 
\newcommand{\by}{\bm{y}} 
 
\newcommand{\iidsim}{\overset{iid}{\sim}}

\newcommand{\transpose}{\mathrm{T}}

\theoremstyle{thmstyleone}%
\theoremstyle{thmstyletwo}%

\theoremstyle{thmstylethree}%

\begin{document}

\title[Comments on: `CliPS: How to identify cluster distributions in Bayesian
mixture models']{Comments on: `CliPS: How to identify cluster distributions in Bayesian
mixture models'}


\author*[1]{\fnm{Sara} \sur{Wade}}\email{sara.wade@ed.ac.uk}



\affil[1]{\orgdiv{School of Mathematics and the Maxwell Institute}, \orgname{University of Edinburgh}, \orgaddress{\street{James Clerk Maxwell Building, Mayfield Rd}, \postcode{EH9 3FD}, \state{Edinburgh}, \country{United Kingdom}}}




\abstract{I congratulate the authors on their contribution, which addresses important challenges in understanding posterior distributions in Bayesian mixture models. As underlined by the authors, the Bayesian framework has several advantages, yet label-switching, identifiability issues, and the trans-dimensional nature of parameter spaces make exploiting the Bayesian approach and  understanding the resulting posterior difficult in practice. The paper offers clear explanations and a simple, intuitive procedure to address this by clustering the Markov chain Monte Carlo (MCMC) draws of component-specific parameters. It is clearly illustrated on several examples.}

\maketitle

\section{Main contribution}\label{sec:contribution}

The authors \citep{malsiner2026clips} develop a procedure to make inference about the underlying sub-populations present in the population based on Markov chain Monte Carlo (MCMC) draws of component-specific parameters from Bayesian mixture models. For data $\by = (\by_1, \ldots, \by_N)$, the mixture model assumes:
\begin{align*}
    \by_i \mid \bTheta_K, \bmeta_K \sim \sum_{k=1}^K \eta_k f(\by_i \mid \btheta_k),
\end{align*}
where the component weights $\bmeta_K = (\eta_1,\ldots, \eta_K)$ are non-negative and sum to one and the $k$-th component-distribution $f(\cdot \mid \btheta_k)$ is parametrized by the $d$-dimensional vector $\btheta_k = (\theta_{k1},\ldots, \theta_{kd})^\transpose$. This formulation is very general and applied in various domains, such as biology, business, economics, and medicine. The Bayesian model is completed with prior distributions:
\begin{align*}
    \btheta_k \mid \zeta &\iidsim p(\btheta_k \mid \zeta)\\
    \bmeta_K \mid K &\sim \mathcal{D}_K(\gamma_K),
\end{align*}
where $p(\btheta_k \mid \zeta)$ is the prior for the component-specific parameters with hyperparameter $\zeta$ and $\mathcal{D}_K(\gamma_K)$ denotes a symmetric $K$-dimensional Dirichlet distribution. In practice the number of clusters is rarely known, and the authors consider and review various Bayesian approaches to address this. In all approaches, MCMC is commonly used to estimate the Bayesian mixture model, producing a large number of posterior draws of the component specific parameters  $$\left(\btheta_{1:K_+^{(m)}}^{(m)}\right) \text{ for  } m=1,\ldots, M,$$ where $K_+^{(m)}$ denotes the $m$-th draw of the number of clusters in the data, that is, the number of occupied components, which may vary across iterations\footnote{All parameter draws corresponding to empty components have been removed.}. 

To address the trans-dimensionality of the parameter space, in addition to well-known identifiability issues and label-switching, the authors propose Clustering in the Parameter Space (CliPS) \cite{malsiner2026clips}.  The main idea is to:
\begin{enumerate}
    \item Filter the MCMC draws by eliminating all iterations with $K_+^{(m)} \neq \widehat{K}_+$, for a suitable estimate $\widehat{K}_+$, such as the posterior mode. Let $M_{\widehat{K}_+}$ denote the remaining number of iterations. 
    \item Apply clustering, namely $k$-means, to group the $\widehat{K}_+ M_{\widehat{K}_+}$ component-specific parameter draws $\btheta_k^{(m)}$ for $k=1, \ldots, \widehat{K}_+$ and $m = 1,\ldots, M_{\widehat{K}_+}$. This produces group assignments $I_k^{(m)}$ for the $k$-th component of the $m$-th iteration.  
    \item Lastly, the iterations are filtered once again and reordered, by first removing all those whose group assignments $\rho_m = (I_1^{(m)}, \ldots, I_{\widehat{K}_+}^{(m)})$ do not define a permutation of $(1,\ldots,\widehat{K}_+)$. This ensures that every component-specific parameter in the $m$-th draw is associated to a unique cluster. The relative number of retained partitions provides a diagnostic indicating how well-separated the clusters are. A unique labeling of the final set of retained draws is achieved by reordering the components through the inverse $\rho_m^{-1}$ of $\rho_m$; these reordered draws characterize the distribution of the component-specific parameters given $\widehat{K}_+$ clusters.  
\end{enumerate}
The approach is elegantly simple and demonstrated through several examples.

\section{Extensions and alternatives to clustering in the parameter space}

A particularly appealing feature of CliPS is that it transforms a difficult relabeling problem into a geometrically intuitive clustering problem in parameter space. This perspective naturally suggests several possible extensions, including alternative clustering strategies that reduce the need for filtering and discarding information  and accommodate more flexible cluster shapes. 
Motivated by the structure visible in Figure 8b of the paper \cite{malsiner2026clips}, one possible alternative is DBSCAN \cite[Density Based Spatial Clustering of Applications with Noise,][]{ester1996density}.  DBSCAN could be applied directly to cluster \textit{all} draws $\btheta_k^{(m)}$ corresponding to non-empty components. Since DBSCAN does not force every draw to belong to a group, it can automatically filter low-density draws (e.g. corresponding to a small cluster that is occasionally split off). In this setting, the group assignment is $I_k^{(m)} \in \lbrace 0,1, \ldots, \widehat{K}_{+} \rbrace$, with $0$ denoting noise and $\widehat{K}_{+}$ denoting the number of clusters identified by DBSCAN. Thus, $\rho_m = (I_1^{(m)}, \ldots, I_{K_+^{(m)}}^{(m)})$ defines a sub-partition. Filtering draws that consist of duplicated non-noise cluster labels, e.g. $\rho_m = (0,1,0,2,3,2)$, is still sensible, but draws that consist of unique subsets of non-noise labels can be retained, e.g. $\rho_m = (0,1,0,0,3,0)$. This provides a unique ordering across draws of potentially varying numbers of clusters. Moreover, DBSCAN allows more general shapes in the cluster distributions.

\begin{figure}[!t]
    \centering
    \subfigure[CliPS: $k$-means, $\widehat{K}_+=4$]{\includegraphics[width=0.49\linewidth]{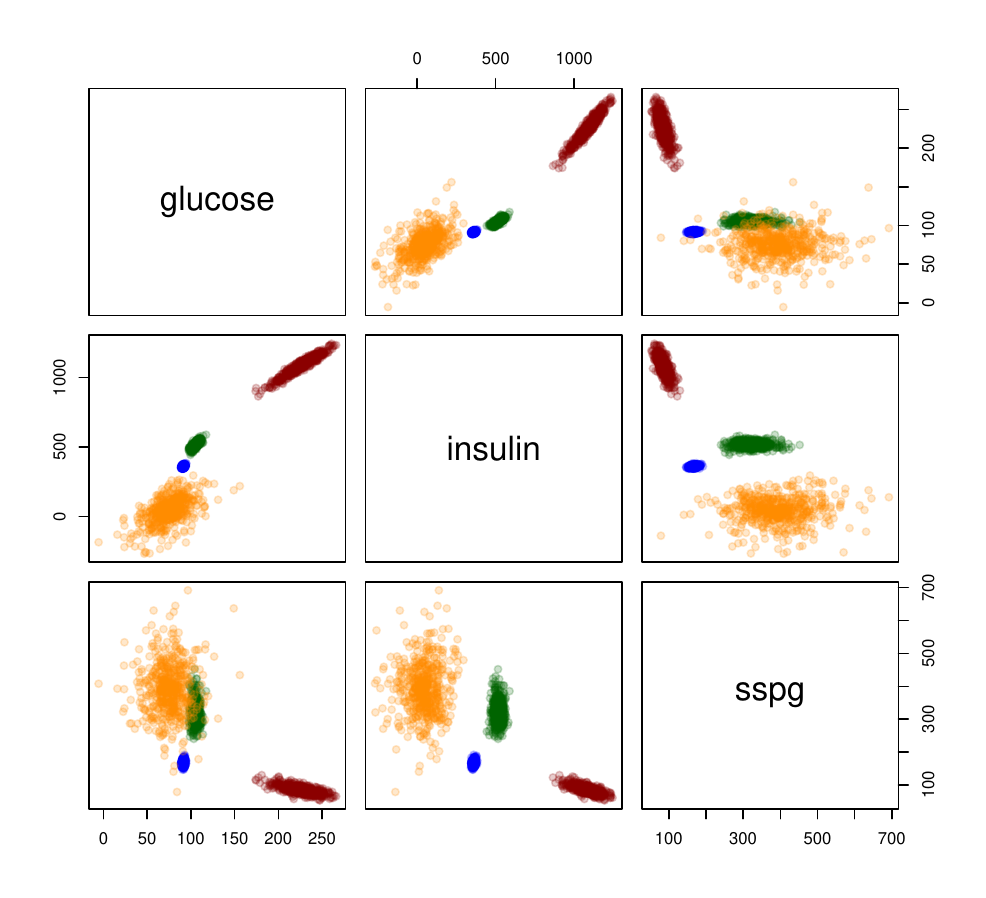}\label{fig:diabetes_kmeans}}
    \subfigure[CliPS: DBSCAN]{\includegraphics[width=0.49\linewidth]{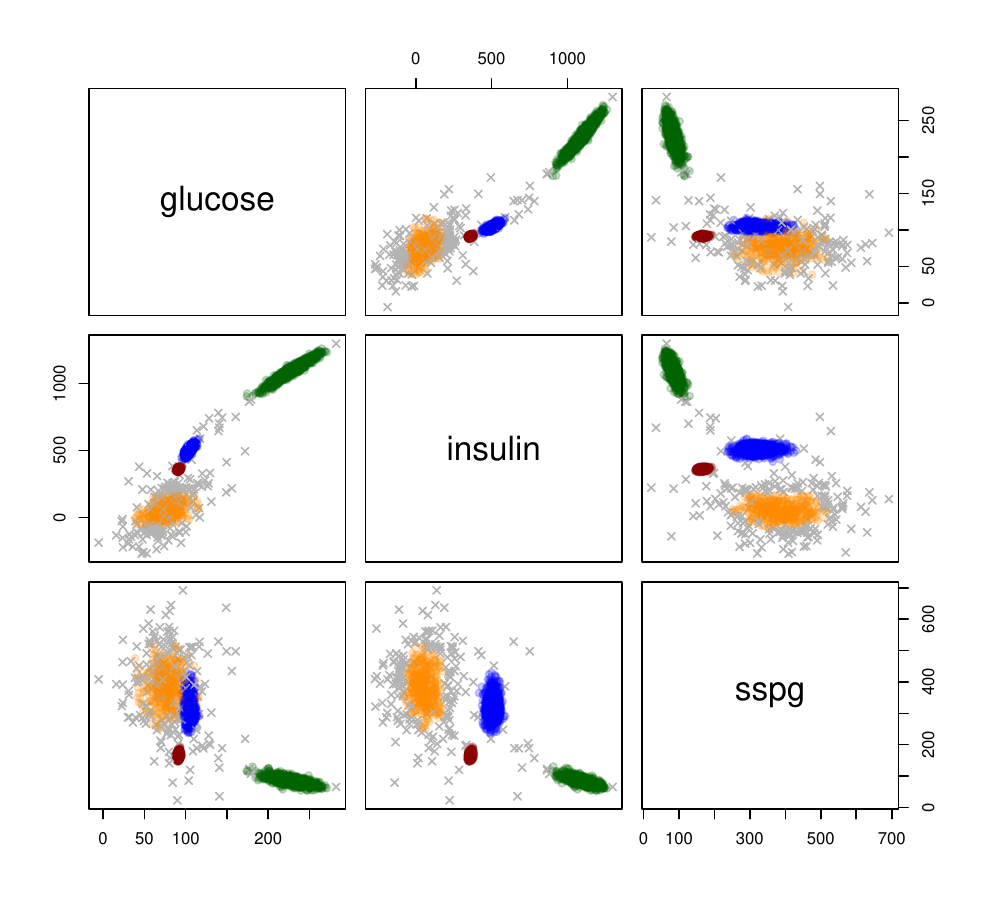}\label{fig:diabetes_dbscan}}\\
    \subfigure[Posterior of $K_{+}$]{\includegraphics[width=0.49\linewidth]{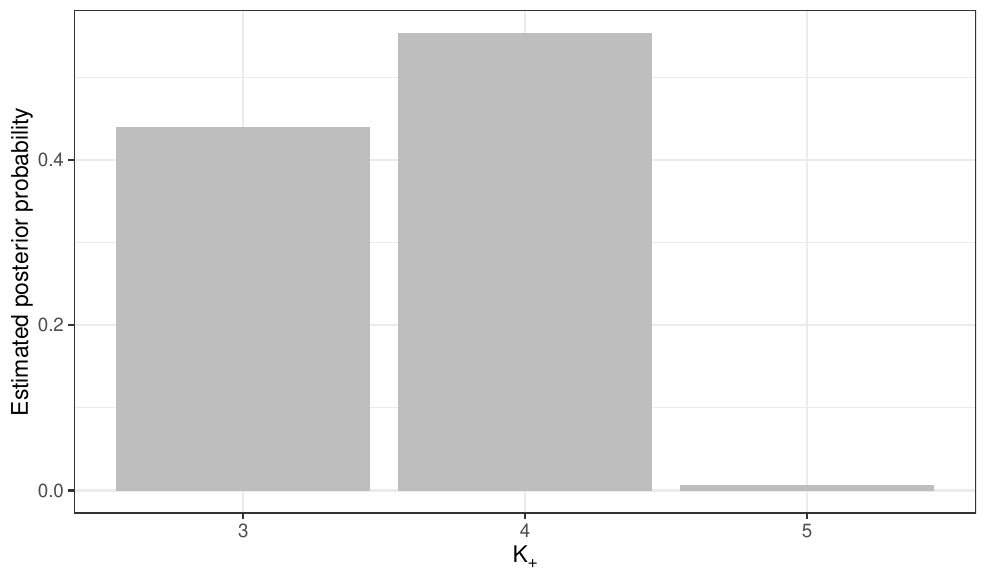}\label{fid:diabetes_k_post}}
     \subfigure[DBSCAN cluster reliability]{\includegraphics[width=0.49\linewidth]{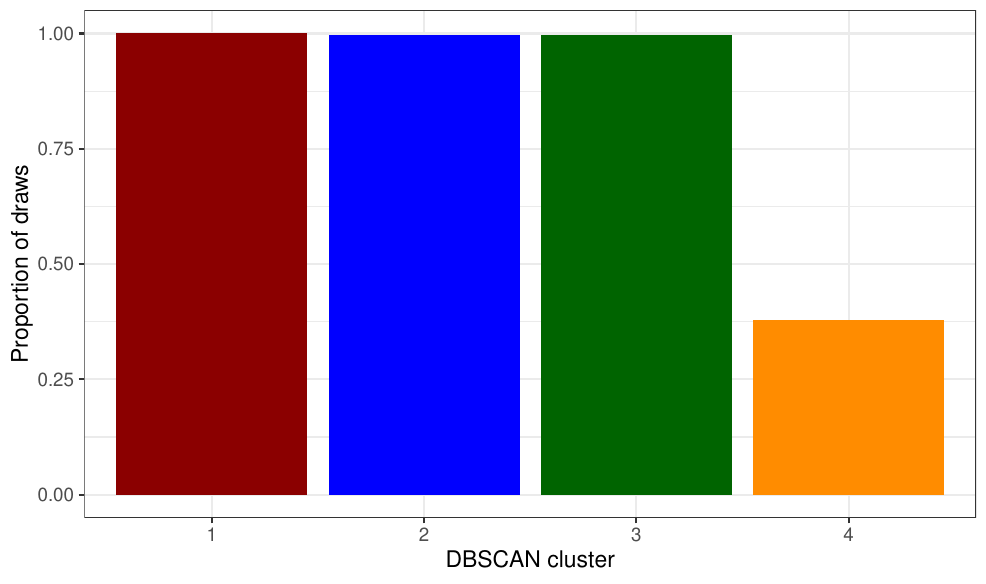}\label{fig:diabetes_dbscan_cluster}}
    \caption{Diabetes example: summary of the cluster distributions based on MCMC draws of the component-specific means from a static Bayesian mixture. Top left: CliPS with $\widehat{K}_+=4$ at posterior mode retains 518 of the 1,000 initial MCMC draws (the non-permutation rate is $0.0633$ and $p(K_+=4 | \by) = 0.553$). Top right: DBSCAN retains 999 of the 1,000 initial MCMC draws (the non-permutation rate is $0.032$). Draws are colored by the DBSCAN cluster labels, with those identified as noise marked with a gray cross. Bottom left: posterior on the number of clusters. Bottom right: Proportion of draws in DBSCAN in which the cluster is present. This reflects the reliability of the cluster, since a value of 1 indicates it is present in all MCMC draws.  }
    \label{fig:diabetes}
\end{figure}

As a simple illustration, I reran the \texttt{diabetes} data set example, available through the \texttt{R} package \texttt{mclust} \citep{scrucca2023model}.  
Prior specification follows  \cite{malsiner2026clips}, except the use of a static mixture\footnote{Due to technical difficulties in running the dynamic mixture in \texttt{telescope}, the static version is implemented following the tutorial, with a prior $e_0 \sim \text{Gam}(1,20)$.} within the \texttt{telescope} package to obtain MCMC draws \citep{fruhwirth2021generalized}.  In this case, the posterior mode of $K_+$ is 4, and after filtering for $K_{+}^{(m)}=4$, the number of MCMC draws retained is 553 out of $1,000$. Applying CliPS, the non-permutation is $0.0633$, and the cluster distributions, visualized in Figure~\ref{fig:diabetes_kmeans}, are from 518 draws. In contrast, I applied DBSCAN to all MCMC draws (with \texttt{minPts}$=10$ and \texttt{eps}$=30$ chosen based on the kNN distance plot). Cluster distributions colored by DBSCAN group assignments are shown in Figure~\ref{fig:diabetes_dbscan}, with noise assignments marked with a gray cross. This is based on 999 of the initial 1,000 MCMC draws (only one is removed after filtering draws which contain duplicated clusters). Moreover, for each DBSCAN group, we can assess its reliability by the proportion of retained draws in which it is contained. High values for groups 1, 2, and 3 indicate that they are present in (almost) all iterations, whereas there is more uncertainty in the presence of group 4.

\section{Comparison with other approaches}

The paper \cite{malsiner2026clips} provides an overview in the Introduction of other approaches to summarize posterior distributions in Bayesian mixture models, but does not provide empirical comparisons in the examples. Thus, to further contextualize the contribution of CliPS, I compare summaries of related approaches on selected examples.  In particular, while the paper \cite{malsiner2026clips} emphasizes the  `use the data to infer a statistical model which allows to determine
the cluster distributions of the data generating process and thus characterizes the
population of interest, beyond focusing only on the partition of the sample at hand',  
the subsequent summarization step yields an estimated partition induced by the identified cluster distributions. In contrast, other approaches first estimate the partition \citep[e.g.][]{Lau07,Fritsch09,wade2018bayesian,rastelli2018optimal,dahl2022search} and subsequently, obtain inference on the sub-population parameters conditioned on the partition estimate. Therefore, it is natural to  compare such approaches. In addition, while not directly compared here, we also highlight the elegant work of \cite{nguyen2026summarizing}, which is more directly related to CliPS. Using a decision-theoretic framework, \cite{nguyen2026summarizing}
identify a point estimate of the underlying mixing measure by minimizing posterior expected sliced Wasserstein distance, and subsequently obtain a partition estimate given the estimated mixing measure.

\begin{figure}[!t]
    \centering
    \subfigure[Partition minimizing the VI lower bound]{\includegraphics[width=0.49\linewidth]{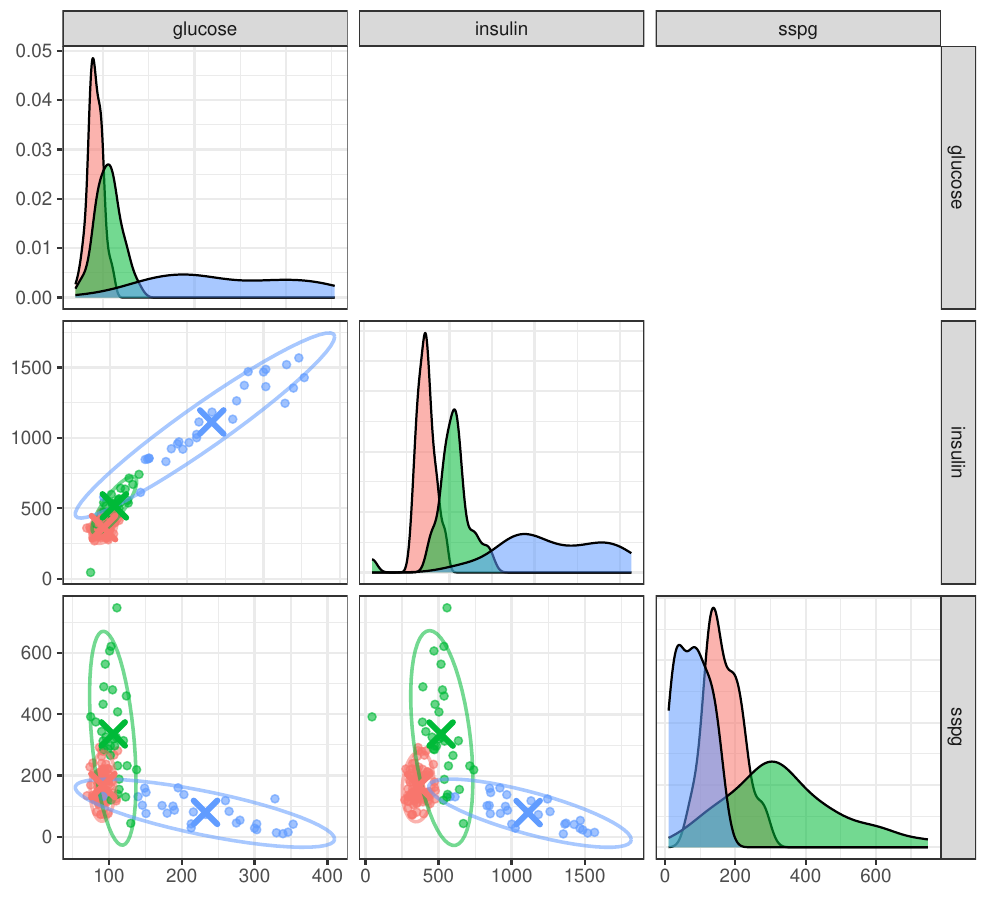}\label{fig:diabetes_minvilb}}
    \subfigure[Lower vertical bound]{\includegraphics[width=0.49\linewidth]{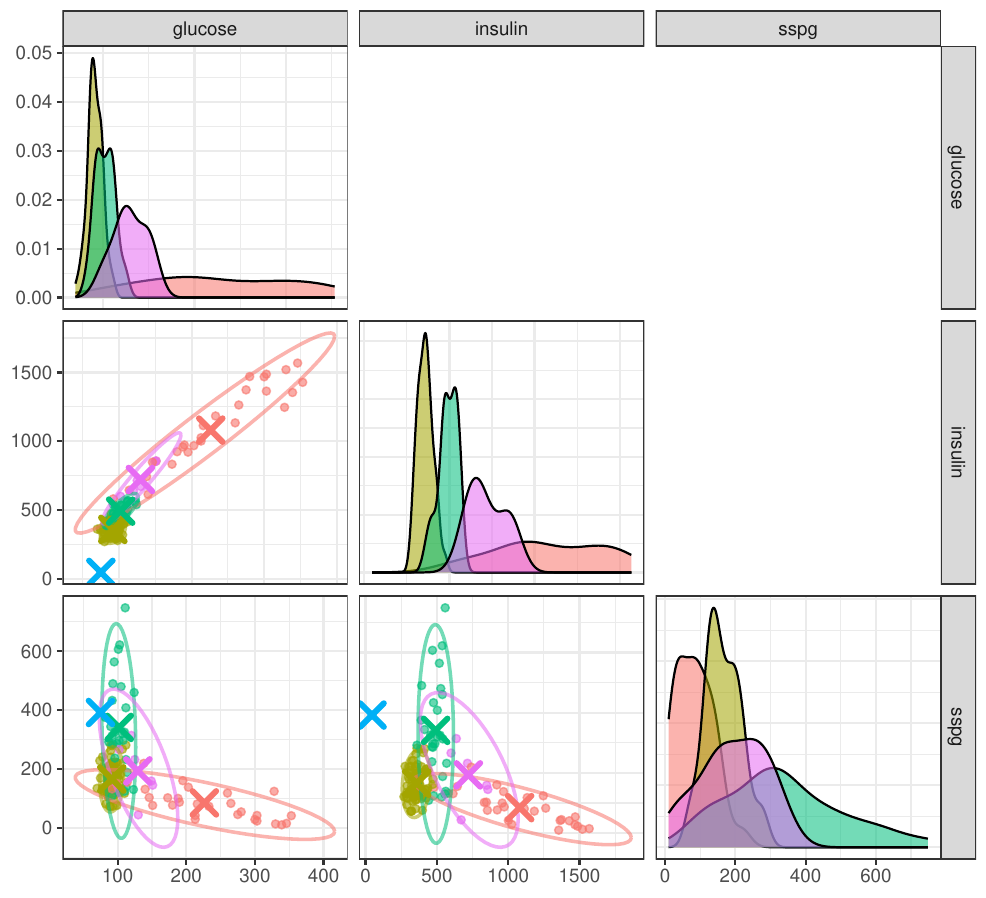}\label{fig:diabetes_lower}}
    \caption{Diabetes example: summary of the posterior partition structure based on MCMC draws from a static Bayesian mixture. Pair plots of the observed data colored by cluster, with the cross and ellipse illustrating the fitted mean and covariance conditioned on the estimated partition. Kernel density estimates are shown on the diagonal. Left: the partition  minimizing the posterior expected VI lower bound \cite{wade2018bayesian} contains 3 clusters. Right:  the lower bound in the $95\%$ credible ball around the estimate contains five clusters.   }
    \label{fig:diabetes_minvi}
\end{figure}

\begin{figure}[!t]
    \centering
    \subfigure[WASABI: Particle 1]{\includegraphics[width=0.49\linewidth]{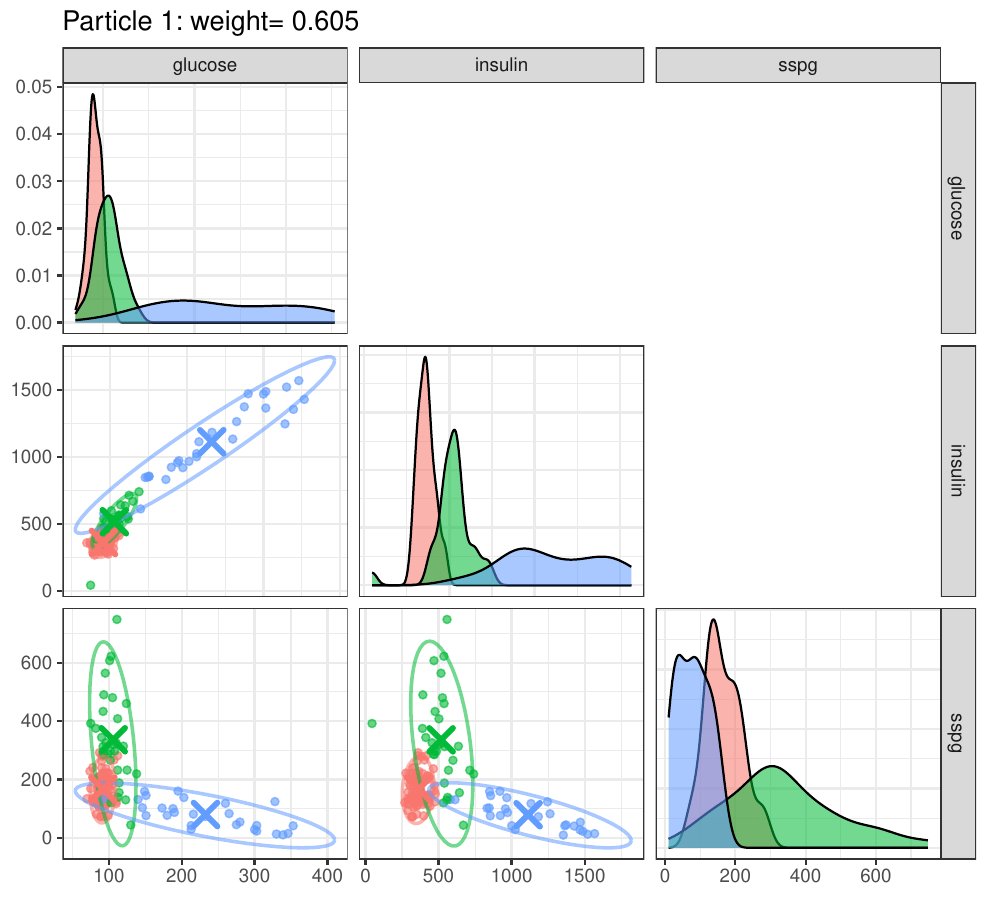}\label{fig:diabetes_wasabi_p1}}
    \subfigure[WASABI: Particle 2]{\includegraphics[width=0.49\linewidth]{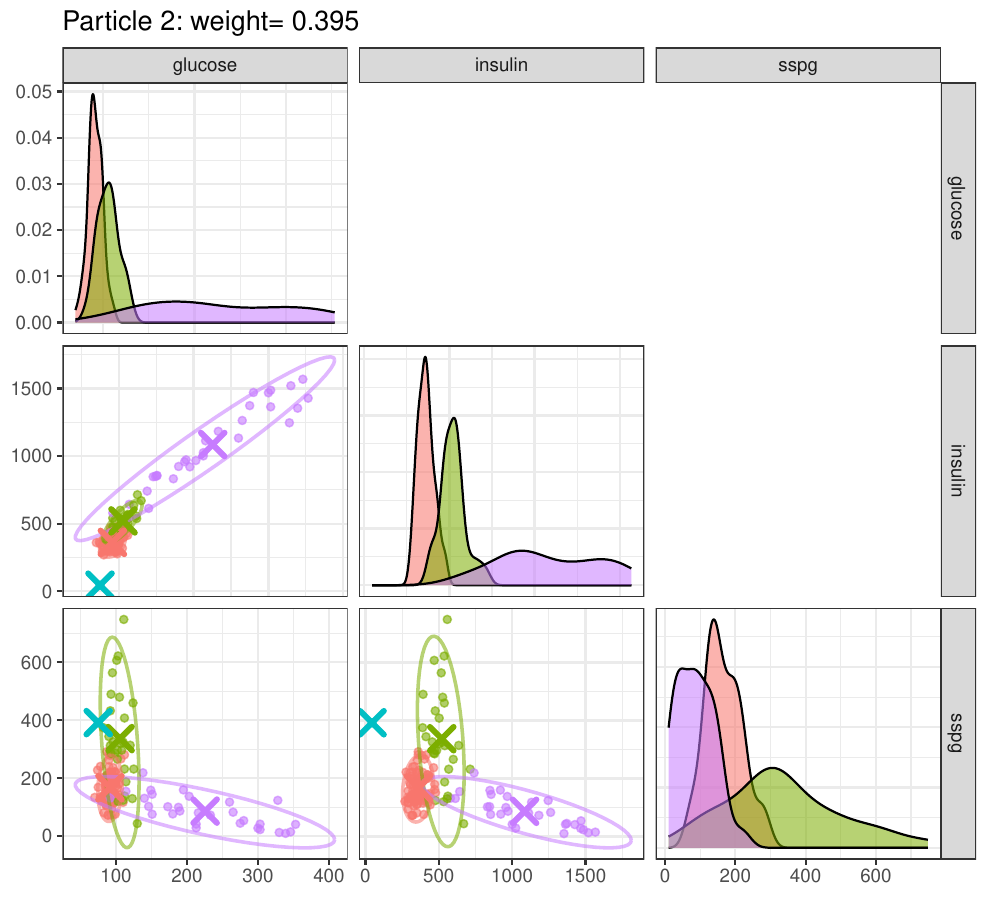}\label{fig:diabetes_wasabi_p2}}
    \caption{Diabetes example: summary of the posterior partition structure based on MCMC draws from a static Bayesian mixture. Pair plots of the observed data colored by cluster, with the cross and ellipse illustrating the fitted mean and covariance conditioned on the estimated partition. Kernel density estimates are shown on the diagonal. WASABI \citep{balocchi2025understanding} summarizes the posterior with a small number of particles (i.e. partitions) capturing different modes in the posterior. Left: the first particle  contains 3 clusters and has a weight of $0.605$. Right: the second particle  contains 4 clusters and has a weight of $0.395$.   }
    \label{fig:diabetes_wassabi}
\end{figure}

We illustrate these comparisons using the \texttt{diabetes} example, comparing the CliPS summary with that obtained from \cite{wade2018bayesian} and \cite{balocchi2025understanding}. In \cite{wade2018bayesian}, an estimate of partition is obtained by minimizing the posterior expected variation of information \citep[VI,][]{Meila07} and credible balls are constructed around this estimate. Figure~\ref{fig:diabetes_minvilb} shows the partition estimate containing 3 clusters, along with the fitted mean and covariance given the partition. The 95\% credible ball around this estimate can be summarized by the upper vertical bound (most distance among those in the credible ball with the fewest number of clusters),  the lower  vertical bound (most distance among those in the credible ball with the greatest number of  clusters), and the horizontal bound (most distance among those in the credible ball). For conciseness, we show only the lower vertical bound in Figure~\ref{fig:diabetes_lower} containing five clusters, one of which is a singleton and another with only 6 observations, and note that the upper vertical bound has three clusters, while the horizontal bound has four. Comparing with CliPS in Figure~\ref{fig:diabetes_kmeans}, the main conclusions are similar. However, 
CliPS places greater emphasis on the possibility of a small fourth sub-population. 
Instead, CliPS-DBSCAN (Figure~\ref{fig:diabetes_dbscan}) estimates a reliability probability of less than 0.5 for this fourth, small cluster, and this cluster is not present in the partition estimate in Figure~\ref{fig:diabetes_minvilb}, although is present in the 95\% credible ball.

\cite{balocchi2025understanding} propose to compress the posterior on the partition space with a small number of weighted partition estimates reflecting the main clustering patterns in the posterior. This compressed posterior places the highest approximate posterior weight of $0.605$ on the partition with three clusters in Figure~\ref{fig:diabetes_wasabi_p1},  and a smaller, yet still substantial weight on the partition with four clusters Figure~\ref{fig:diabetes_wasabi_p2}. Interestingly, this closely resembles the posterior summary provided by CliPS-DBSCAN in Figure~\ref{fig:diabetes_dbscan}. 

Next, we consider the simulated example in \cite{balocchi2025understanding}, consisting of a two-dimensional Gaussian mixture with four components, each with mean
$(\pm m, \pm m)$ located in one of the four quadrants and identity covariance matrix. Here $m=1.25$, so that the clusters are only mildly separated. The authors mention that CliPS is appropriate for well separated clusters. Since the degree of cluster separation is rarely known in advance, it is also interesting to investigate how CliPS behaves in more challenging settings where the underlying structure is only weakly present.

\begin{figure}[!t]
     \centering
     \subfigure[Posterior of $K_+$]{\includegraphics[width=0.45\linewidth]{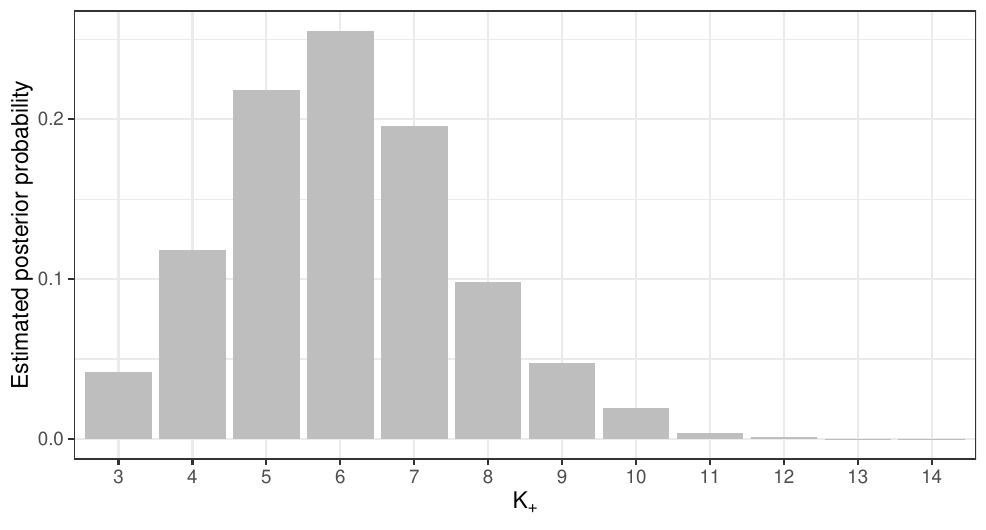}\label{fig:4modes_k}}
          \subfigure[CliPS: $\widehat{K}_+=6$]{\includegraphics[width=0.4\linewidth]{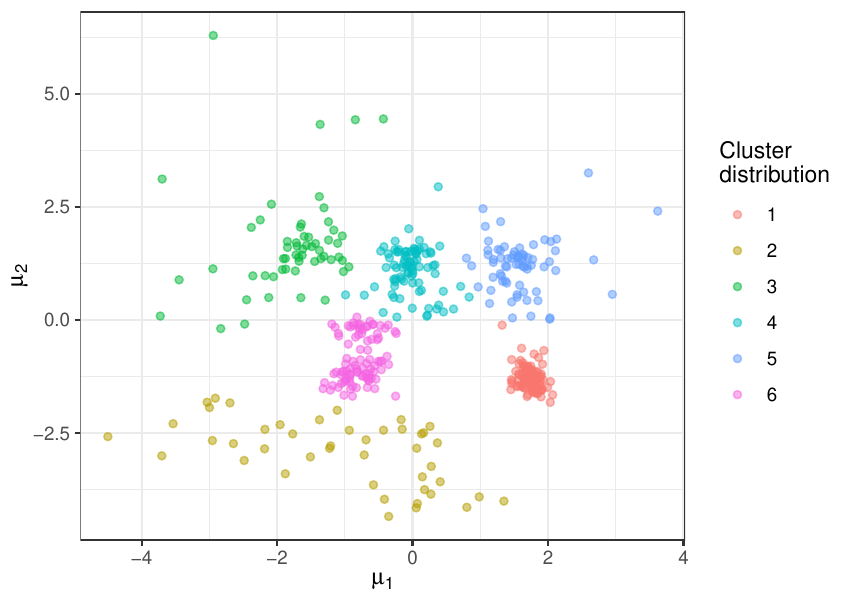}\label{fig:4modes_clips}}\\
         \subfigure[WASABI: 2 particles]{\includegraphics[width=0.65\linewidth]{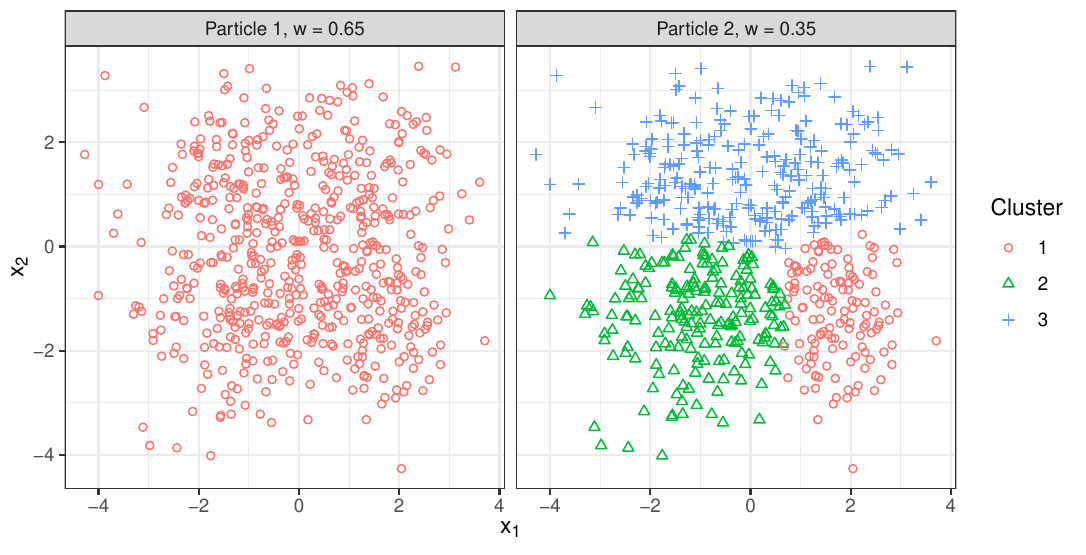}\label{fig:4modes_wasabi}} 
     \caption{Slightly multi-modal example: $n=600$ observations are generated from a two-dimensional Gaussian mixture with four components, each with mean
$(\pm 1.25, \pm 1.25)$ located in one of the four quadrants and identity covariance matrix. MCMC draws are obtained by fitting a DPM of Gaussian. Top left: posterior on the number of clusters. Top right: CliPS with $\widehat{K}_+ = 6$ at the posterior mode retains only 93 of the initial 10,000 MCMC draws (the non-permutation rate is 0.964 and $p(K_+=6 | \by) = 0.255$). Bottom: WASABI summary with two particles, the first consists of a single cluster with a weight of $0.65$, while the second contains 3 clusters with a weight of $0.35$. }
     \label{fig:4modes}
 \end{figure}

MCMC draws are obtained by fitting a diagonal
location-scale Dirichlet process mixture (DPM) \citep{JSSv100i15}. The posterior on the number of clusters (Figure~\ref{fig:4modes_k}) is concentrated around the range of 3 to 10 clusters. In CliPS, we set $\widehat{K}_+=6$ to the posterior mode. However, the non-permutation rate is high, and the CliPS summary provided in Figure~\ref{fig:4modes_clips}, which is based on only 93 of the original $10,000$ MCMC draws, provides a less informative summary of the true component-specific means at  $(\pm 1.25, \pm 1.25)$. As the authors suggest, when the non-permutation rate is high (as in this example), CliPS is not suitable and other summarization methods may be more appropriate (such as the WASABI summary in Figure~\ref{fig:4modes_wasabi}). But, this raises an interesting question: can we quantify how high the non-permutation rate should be for CliPS to provide an effective summary? Moreover, while not shown here for brevity, increasing the separation (e.g. $m=2$) substantially reduces the ambiguity in the posterior summary, increasing the non-permutation rate and making CliPS more effective. This raises a related interesting question for future work: can one characterize the degree of cluster separation required for CliPS to provide an effective summary?

\section{Further comments}
I end with a few further comments.
\begin{enumerate}
\item What distribution do the retained, reordered draws characterize? The approach and resulting interpretations could be further strengthened by more formally characterizing the distribution that is approximated by the retained draws. The first filtering clearly retains  draws that approximate the posterior distribution of parameters conditioned on $\widehat{K}_+$ clusters, but it is not as clear to understand how the final filtering step affects the target posterior distribution that is being approximated. A more formal characterization would help clarify the inferential target of the procedure and facilitate interpretation of the resulting cluster distributions.  
    \item The authors further propose to use a lower-dimensional functional $\varphi(\btheta_k)$ of the component-specific parameters, and apply $k$-means in this lower-dimensional space. They provide examples, such as clustering only the means $\bmu_k$ and not the covariance matrix $\bSigma_k$, which may be considered nuisance parameters in mixtures of multivariate Gaussians. In general, choosing an appropriate function is not straightforward, but it may also help to combat the curse of dimensionality in high-dimensional parameter spaces. In genomics, and other fields where high-dimensions are a concern, standard pipelines \cite{gribov2010seurat, wolf2018scanpy} first apply principal component analysis (PCA) and then cluster in the lower-dimensional space spanned by the  selected number of principal components. Similar dimensionality-reduction strategies may prove useful within the CliPS framework, particularly when clustering is performed in high-dimensional parameter spaces where distance-based methods can become less reliable.
\end{enumerate}

\backmatter

\section*{Declarations}

\bmhead{Funding Statement}

S.W. acknowledges funding by the Engineering and Physical Sciences Research Council (EP/Y028783/1) and by a Leverhulme Trust Research Fellowship (RF-2026-207\textbackslash9). 

\bmhead{Competing Interests Statement}

The author declares no competing interests.









\bibliography{main}

\end{document}